\documentclass[article,preprint,groupedaddress]{revtex4}
\usepackage{epsfig}
\usepackage{graphicx}

\begin{document}
\title{Black-hole evaporation, cosmic censorship, and a quantum lower bound on the Bekenstein-Hawking temperature}
\author{Shahar Hod}
\address{The Ruppin Academic Center, Emeq Hefer 40250, Israel}
\address{ }
\address{The Hadassah Institute, Jerusalem 91010, Israel}
\date{\today}

\begin{abstract}
\ \ \ The semi-classical Hawking evaporation process of
Reissner-Nordstr\"om black holes is analyzed. It is shown that this
quantum mechanism may turn a near-extremal black-hole spacetime with
$T_{\text{BH}}\lesssim \hbar^2/G^2M^3$ into an horizonless naked
singularity, thus violating the Penrose cosmic censorship
conjecture. It is therefore conjectured that, within the framework
of a self-consistent quantum theory of gravity, the
Bekenstein-Hawking temperature should be bounded from below by the
simple relation $T_{\text{BH}}\gtrsim \hbar^2/G^2M^3$.
\end{abstract}
\bigskip
\maketitle


\section{Introduction}

Following Bekenstein's proposal that black holes have a well defined
entropy \cite{Bekt}, Hawking, using a semi-classical analysis
\cite{Hawt}, has revealed the intriguing fact that these fundamental
objects of general relativity are characterized by thermally
distributed (filtered black-body) radiation spectra. In particular,
the emission spectra of the canonical family of charged and rotating
Kerr-Newman black holes are characterized by the well defined
Bekenstein-Hawking temperature \cite{Noteunit}
\begin{equation}\label{Eq1}
T_{\text{BH}}={{\hbar(r_+-r_-)}\over{4\pi G(r^2_++a^2)}}\  .
\end{equation}
Here
\begin{equation}\label{Eq2}
r_{\pm}=GM\pm(G^2M^2-GQ^2-a^2)^{1/2}\
\end{equation}
are the (inner and outer) horizon radii which characterize the
black-hole spacetime, where $\{M,Q,a\}$ are respectively the
asymptotically measured black-hole mass, electric charge, and
angular momentum per unit mass.

Inspection of the functional expression (\ref{Eq1}) for
$T_{\text{BH}}=T_{\text{BH}}(M,Q,a)$ immediately reveals the
intriguing fact that the semi-classical Bekenstein-Hawking
temperature approaches zero in the $M^2-Q^2-a^2\to0$ limit of
near-extremal black holes.

In the present paper we shall address the following physically
interesting question: Can the Bekenstein-Hawking temperature of a
given mass black hole be made arbitrarily small? According to Page
\cite{Pageev,Pagepriv}, the answer to this question is `yes'. Page
has based his argument for the non-existence of a lower bound on the
Bekenstein-Hawking temperature on the fact that, for near-extremal
Reissner-Nordstr\"om ($a=0$) black holes in the large-mass regime
\cite{Pageev,Noteelc}
\begin{equation}\label{Eq3}
M\gg{{e\hbar}\over{\pi m^2_e}}\  ,
\end{equation}
the emission of charged massive fields is exponentially suppressed
as compared to the emission of neutral massless (electromagnetic and
gravitational) fields \cite{Pageev,Pagepriv}. It was therefore
argued by Page \cite{Pageev,Pagepriv} that, by emitting neutral
fields which reduce the black-hole mass (without reducing its
electric charge), a charged Reissner-Nordstr\"om black hole can
approach arbitrarily close to the zero-temperature (extremal) limit
$T_{\text{BH}}\to0$ \cite{NoteHawr,Hawr,Hodext}.

The main goal of the present compact paper is to reveal the fact
that near-extremal Reissner-Nordstr\"om black holes in the regime
$T_{\text{BH}}\lesssim \hbar^2/G^2M^3$, if they exist, may violate,
through the Hawking emission process, the black-hole condition
$Q\leq M$ which is imposed by the Penrose cosmic censorship
conjecture \cite{HawPen,Pen}. This fact, to be proved below,
suggests that, in order to guarantee the validity of the fundamental
Penrose cosmic censorship conjecture, the Bekenstein-Hawking
temperature of a given mass black hole should be bounded from below
by the simple relation $T_{\text{BH}}\gtrsim \hbar^2/G^2M^3$.

\section{The semi-classical Hawking evaporation process of
near-extremal Reissner-Nordstr\"om black holes}

In the present section we shall analyze the physical and
mathematical properties which characterize the Hawking radiation
spectra of near-extremal Reissner-Nordstr\"om black holes in the
regime \cite{Notepqm}
\begin{equation}\label{Eq4}
0\leq\Delta\equiv M-Q\ll M\  .
\end{equation}
Here $\Delta$ is the excess energy of the charged
Reissner-Nordstr\"om spacetime above the minimal mass (extremal)
black-hole configuration with $M=M_{\text{min}}(Q)=Q$. As explicitly
shown in \cite{Pageev}, the semi-classical decay of near-extremal
Reissner-Nordstr\"om black holes in the large-mass regime
(\ref{Eq3}) is dominated by the emission of massless neutral photons
with unit angular momentum \cite{Noteexs}.

The black-hole radiation power for one bosonic degree of freedom is
given by the semi-classical Hawking integral relation \cite{Page}
\begin{equation}\label{Eq5}
P={{\hbar G}\over{2\pi}}\sum_{l,m}\int_0^{\infty} {{\Gamma
\omega}\over{e^{\hbar\omega/T_{\text{BH}}}-1}}d\omega\  ,
\end{equation}
where $\{l,m\}$ are the angular (spheroidal and axial) harmonic
indices which characterize the emitted field mode, and the
frequency-dependent dimensionless parameters
$\Gamma=\Gamma_{lm}(\omega)$ are the greybody factors which
characterize the linearized interaction (scattering) of the field
mode with the curved black-hole spacetime \cite{Page}.

It is interesting to point out that the characteristic thermal
factor $\omega/(e^{\hbar\omega/T_{\text{BH}}}-1)$ that appears in
the semi-classical expression (\ref{Eq5}) for the Hawking radiation
power implies that the black-hole-field emission spectra peak at the
characteristic dimensionless frequency
\begin{equation}\label{Eq6}
{{\hbar\omega^{\text{peak}}}\over{T_{\text{BH}}}}=O(1)\ .
\end{equation}
Since the semi-classical Bekenstein-Hawking temperature (\ref{Eq1})
of a near-extremal Reissner-Nordstr\"om black hole in the regime
(\ref{Eq4}) is characterized by the strong dimensionless inequality
\cite{Noteex}
\begin{equation}\label{Eq7}
{{GMT_{\text{BH}}}\over{\hbar}}\ll1\  ,
\end{equation}
one deduces from (\ref{Eq6}) that, for near-extremal black holes in
the regime (\ref{Eq7}), the characteristic field frequencies which
constitute the Hawking black-hole emission spectra are characterized
by the strong dimensionless inequality
\begin{equation}\label{Eq8}
M\omega^{\text{peak}}\ll1\  .
\end{equation}

Interestingly, and most importantly for our analysis, it has been
demonstrated in \cite{Pageev} that, in the low-frequency regime
(\ref{Eq8}) which dominates the emission spectra of the
near-extremal black holes, the dimensionless greybody factors
$\Gamma_{lm}(\omega)$ that appear in the integral relation
(\ref{Eq5}) can be expressed in a closed analytic form. In
particular, one finds the simple expression \cite{Pageev}
\begin{equation}\label{Eq9}
\Gamma_{1m}={1\over9}\epsilon^8\nu^4(1+\nu^2)(1+4\nu^2)\cdot[1+O(M\omega)]\
\end{equation}
for the characteristic greybody factor of unit ($l=1$) angular
momentum photons \cite{Notedom}, where here we have used the
dimensionless physical quantities
\begin{equation}\label{Eq10}
\epsilon\equiv {{r_+-r_-}\over{r_+}}\ \ \ \ \text{and}\ \ \ \
\nu\equiv {{\hbar\omega}\over{4\pi GT_{\text{BH}}}}\  .
\end{equation}

Substituting Eq. (\ref{Eq9}) into Eq. (\ref{Eq5}), and using the
relations [see Eqs. (\ref{Eq1}), (\ref{Eq4}), and (\ref{Eq10})]
\begin{equation}\label{Eq11}
\epsilon=\sqrt{{{8\Delta}\over{M}}}\cdot[1+O(\epsilon)]\ \ \ \
\text{and}\ \ \ \
\omega=\nu\sqrt{{{8\Delta}\over{M^3}}}\cdot[1+O(\epsilon)]\  ,
\end{equation}
one obtains the simple expression \cite{Pageev}
\begin{equation}\label{Eq12}
P={{\hbar\epsilon^{10}}\over{3\pi GM^2}}\int_0^{\infty} {\cal
F}(\nu)d\nu\ \ \ \ \text{with}\ \ \ \ {\cal
F}(\nu)\equiv{{4\nu^9+5\nu^7+\nu^5}\over{e^{4\pi\nu}-1}}
\end{equation}
for the semi-classical Hawking radiation power of the near-extremal
Reissner-Nordstr\"om black holes.

From Eq. (\ref{Eq12}) one finds that the function ${\cal F}(\nu)$,
which determines the energy distribution of the radiated field
modes, has a maximum at
\begin{equation}\label{Eq13}
\nu=\nu_{\text{peak}}\simeq 0.511\  ,
\end{equation}
implying that the characteristic photons in the emission spectra of
the near-extremal black holes have an energy which is of the order
of [see Eqs. (\ref{Eq11}) and (\ref{Eq13})]
\begin{equation}\label{Eq14}
E=\hbar\omega=\hbar
G^{-1}\nu_{\text{peak}}\sqrt{{{8\Delta}\over{M^3}}}\ .
\end{equation}
The Hawking emission of these neutral field modes would reduce the
black-hole mass (without reducing its charge) by $\Delta M=-E$.
Thus, the mass of the resulting black-hole configuration (after the
emission of a characteristic Hawking photon) is given by [see Eqs.
(\ref{Eq4}) and (\ref{Eq14})]
\begin{equation}\label{Eq15}
M_{\text{new}}=M-E=Q+\Delta-E\  .
\end{equation}

Taking cognizance of the Penrose cosmic censorship conjecture
\cite{HawPen,Pen}, one immediately realizes that, in order for the
Hawking evaporation process to respect the black-hole condition
(\ref{Eq4}), the characteristic energy of the emitted photons must
be bounded from above. In particular, using the constraint
\begin{equation}\label{Eq16}
Q\leq M_{\text{new}}\
\end{equation}
on the physical parameters which characterize the black-hole
configuration after the emission of a characteristic Hawking photon,
one finds the lower bound [see Eqs. (\ref{Eq14}), (\ref{Eq15}), and
(\ref{Eq16})]
\begin{equation}\label{Eq17}
\Delta\geq {{8(\hbar\nu_{\text{peak}}/G)^2}\over{M^3}}
\end{equation}
on the excess energy of a given mass Reissner-Nordstr\"om black
hole.

Interestingly, using the relation [see Eqs. (\ref{Eq1}),
(\ref{Eq10}), and (\ref{Eq11})]
\begin{equation}\label{Eq18}
T_{\text{BH}}=\hbar
G^{-1}\sqrt{{{\Delta}\over{2\pi^2M^3}}}\cdot[1+O(\epsilon)]\ ,
\end{equation}
one finds from (\ref{Eq17}) the lower bound
\begin{equation}\label{Eq19}
T_{\text{BH}}\geq {{2\nu_{\text{peak}}\hbar^2}\over{\pi G^2M^3}}\
\end{equation}
on the Bekenstein-Hawking temperature of the near-extremal black
holes. Thus, as opposed to the claim made in \cite{Pageev,Pagepriv},
a near-extremal black hole cannot radiate its excess energy all the
way down to $T_{\text{BH}}\to0$. In particular, our analysis has
revealed the intriguing fact that Reissner-Nordstr\"om black holes
that violate the relation (\ref{Eq19}) would also violate, through
the Hawking evaporation process, the fundamental Penrose cosmic
censorship conjecture \cite{HawPen,Pen}.

\section{Summary and Discussion}

In the present compact paper we have raised the following physically
interesting question: Can the Bekenstein-Hawking temperature of a
given mass black hole be made arbitrarily small? In order to address
this question, we have analyzed the physical and mathematical
properties of the semi-classical Hawking radiation spectra which
characterize near-extremal [see (\ref{Eq4})] Reissner-Nordstr\"om
black holes in the large-mass regime (\ref{Eq3}).

It has been shown that the black-hole evaporation process may turn a
near-extremal black hole with
$\Delta<{{8(\hbar\nu_{\text{peak}}/G)^2}/{M^3}}$ [or equivalently,
with a Bekenstein-Hawking temperature of $T_{\text{BH}}<
{{2\nu_{\text{peak}}\hbar^2}/{\pi G^2M^3}}$] into an horizonless
naked singularity with $Q>M$, thus violating the cosmic censorship
conjecture \cite{HawPen,Pen}.

Our analysis therefore suggests that one of the following physical
scenarios should be valid in a self-consistent quantum theory of
gravity:

(1) The Penrose cosmic censorship conjecture can be violated within
the framework of a quantum theory of gravity.

(2) The Hawking semi-classical treatment of the black-hole
evaporation process breaks down in the regime
$T_{\text{BH}}\lesssim\hbar^2/G^2M^3$ of near-extremal black holes
in such a way that the emission of charged massive fields (which
reduce the black-hole charge and thus increase the black-hole
temperature) dominates over the emission of neutral massless fields.

(3) Near-extremal black holes with $\Delta\lesssim\hbar^2/G^2M^3$ do
not exist within the framework of a self-consistent quantum theory
of gravity. If true, this last scenario implies that the
Bekenstein-Hawking temperature of quantized black holes is bounded
from below by the simple relation \cite{Notealu,Noteind}
\begin{equation}\label{Eq20}
T_{\text{BH}}\gtrsim {{\hbar^2}\over{G^2M^3}}\  .
\end{equation}

\bigskip
\noindent
{\bf ACKNOWLEDGMENTS}
\bigskip

This research is supported by the Carmel Science Foundation. I thank
Don Page for interesting correspondence. I would also like to thank
Yael Oren, Arbel M. Ongo, Ayelet B. Lata, and Alona B. Tea for
stimulating discussions.



\begin{thebibliography}{99}
\bibitem{Noteunit} We shall use natural units in which
$c=k_{\text{B}}=1$.
\bibitem{Noteelc} Here $e$ and $m_e$ are respectively the electric charge and proper
mass of the positron.
\bibitem{NoteHawr} It is worth noting that it has been claimed in the physics literature that
extremal black holes possess some peculiar physical properties. For
instance, it was argued that, despite having non-zero surface areas,
the entropy of extremal black holes is zero \cite{Hawr} (see also
\cite{Hodext} and references therein).
\bibitem{Noteexs} In particular, it has been shown in
\cite{Pageev,Pagepriv} that, in the large-mass regime (\ref{Eq3}),
the black-hole emission of charged massive fields is exponentially
suppressed as compared to the Hawking emission of neutral massless
fields. In addition, for the near-extremal black holes (\ref{Eq4}),
the emission of electromagnetic and gravitational field quanta with
larger values of the angular momentum parameter ($l>1$) is
suppressed, as compared to the emission of photons with unit angular
momentum ($l=1$), by several factors of the large dimensionless
ratio $M/\Delta$ \cite{Pageev,Pagepriv}.
\bibitem{Noteex} Note that near-extremal Reissner-Nordstr\"om black holes are
characterized by the strong inequality $(r_+-r_-)/r_+\ll1$. This
inequality implies the dimensionless relation
$GMT_{\text{BH}}/\hbar\ll1$ [see Eq. (\ref{Eq1})] for the
near-extremal black holes.
\bibitem{Notedom} It is worth emphasizing again that, as explicitly shown in \cite{Pageev},
these unit angular momentum photons dominate the Hawking radiation
spectra of the near-extremal Reissner-Nordstr\"om black holes in the
large-mass regime (\ref{Eq3}).
\bibitem{Notealu} Note that the dimensionless factor
$2\nu_{\text{peak}}/\pi$ in Eq. (\ref{Eq19}) is of order unity.
\end{thebibliography}
\end{document}